\documentclass[twocolumn,preprintnumbers,amsmath,amssymb]{revtex4}

\usepackage{graphicx}  
\begin{document}
\title{
Graphene vertical  hot-electron terahertz  detectors
}
\author{V.~Ryzhii\footnote{Electronic mail: v-ryzhii(at)riec.tohoku.ac.jp}$^{1,2}$, A. Satou$^1$, T.~Otsuji$^{1}$, M.~Ryzhii$^{3}$, 
 V. Mitin$^4$, and M.S. Shur$^5$
}
\affiliation{
$^1$Research Institute for Electrical Communication, Tohoku University, Sendai 980-8577, Japan\\
$^2$ Center for Photonics and Infrared Engineering, 
Bauman Moscow State Technical University 
and Institute of Ultra High Frequency Semiconductor Electronics, Russian Academy of Sciences,
Moscow 111005, Russia\\ 
$^3$ Department of Computer Science and Engineering, University of Aizu, Aizu-Wakamatsu 965-8580, Japan\\
$^4$ Department of Electrical  Engineering, University at Buffalo, Buffalo, NY 1460-1920, USA\\
$^5$ Departments of Electrical, Electronics, and Systems Engineering and Physics, Applied Physics, and Astronomy, Rensselaer Polytechnic Institute, Troy, NY 12180, USA
}

\begin{abstract}
We propose and analyze the  concept of the vertical hot-electron terahertz (THz)  graphene-layer detectors (GLDs) 
based
on  the double-GL  and  multiple-GL structures with the barrier layers made of materials with a moderate 
conduction band off-set (such as tungsten disulfide and related materials).
The operation of these  detectors is enabled by the thermionic emissions from the GLs enhanced by 
the electrons heated 
 by incoming THz radiation. Hence, 
 Hence, these detectors are the hot-electron bolometric detectors. 
The   electron heating is primarily associated with  the intraband absorption (the Drude absorption).
In the frame of the   developed model,  we calculate the  responsivity and detectivity
as functions of the photon energy, GL doping, and the applied voltage for the GL detectors (GLDs) 
 with different number of GLs. 
The detectors based on the cascade multiple-GL structures can exhibit a substantial photoelectric 
gain resulting in the elevated responsivity and detectivity.
The advantages of the THz detectors under consideration are associated with their high sensitivity 
to the normal incident radiation and efficient operation at  room temperature
at the low end of the THz frequency  range. Such GLDs with a metal grating, supporting  the excitation 
of plasma oscillations in the GL-structures by the incident THz radiation, can exhibit a strong resonant 
response at the frequencies of several THz (in the range, where the operation
of the conventional detectors  based on A$_3$B$_5$ materials, in particular THz quantum-well detectors,  
is hindered due to a strong optical phonon radiation absorption  in such materials).
We also evaluate  also the characteristics of GLDs in the mid- and far-infrared ranges where the
electron heating is due to the interband absorption in GLs.
\end{abstract}

\maketitle
\newpage

\newpage
\section{Introduction}

The gapless energy spectrum of graphene~\cite{1} enables using single- or multiple graphene-layer (GL) 
structures for different terahertz (THz) and infrared (IR) photodetectors based on
involving the interband transitions~\cite{1,2,3,4,5,6,7} (see, also~ Refs~\cite{8,9,10,11,12,13,14,15,16,17,18}), where different THz and IR photodetectors based on GLs were 
explored). 
The interband photodetectors use  either the GLs serving as photoconductors or 
the lateral p-i-n junctions. In the latter case, the electrons and holes are generated in the depleted i-region 
and move  to the opposite GL contacts driven by the electric field in the depletion region~\cite{3}. 
The multiple-GL structures with the lateral p-i-n junctions can consist
of either several non-Bernal stacked twisted) GLs as in Ref.~\cite{3} or GLs separated by the barrier layers 
such as thin layers of Boron Nitride (hBN),
Tungsten Disulfide (WS$_2$), or similar materials. 
Such heterostructures 
have recently attracted a considerable interest and enabled several novel devices being proposed and realized~\cite{19,20,21,22,23,24,25,26,27,28,29,30,31}.
The  GL-photodetectors, especially those based on the multiple-GL structures,  
can combine a high responsivity with a relatively low dark current at elevated temperatures (up to room temperatures).
This is because the dark current in the photodetectors in question is mainly determined by the absorption of the optical phonons. 
Since the optical phonon
energy $\hbar\omega_0$ in GLs is rather large (about 0.2~eV), the number of optical phonons is small even at the room temperature. This results in a low  thermal generation rate.
The mechanisms of the thermal generation associated with the absorption of the acoustic phonons
and the Auger processes are forbidden due to the features of the GL energy spectrum. 
However, the interband tunneling in strong lateral electric fields in the i-region can lead to an enhanced generation of the electron-hole pairs and an elevated dark current limiting the photodetector detectivity~\cite{4}.   
Effective THz detection can be achieved in the lateral diodes with  the absorbing GL source and drain sections  separated by an array of grapnene nanoribbons (GNRs),  which  form the potential barriers for hot electrons injected from the source to the drain~\cite{22}.
As shown in this paper, an effective  THz detection
can be achieved 
 in the photodetectors based on double-GL and cascade multiple-GL structures   with the vertical transport of hot electrons over  the barrier layers.
We propose and evaluate such THz  detectors operating in the regime of the thermionic emission of hot electrons from GLs and their vertical transport over the barrier layers.
The advantages of the THz detectors under consideration 
include 
high responsivity and detectivity in a wide spectral range at room temperature
and a relatively high-speed operation.

The paper is organized as follows.
In Sec. II, we discuss the device structures under consideration and the GLD operation principle.  
Section III deals with general formulas for the dark current and photocurrent associated with
the thermionic emission of electrons from GL and controlled by their capture into GLs.
In Sec.~IV, we calculate the variations of the electron temperature in GLs cause by the intraband (Drude) absorption of the incident THz radiation.
In Sections~V and VI, using the formulas obtained  in Sections~III and IV, we derive 
the expressions for the GLD responsivity and dark-current-limited detectivity, respectively.
 In Sec. VII, we discuss how   the electron capture in the GLs affects
the GLD responsivity and detectivity.
In Sec.~VIII, we consider the possibility to use the plasmonic resonances and get 
an enhanced response at elevated frequencies.
Section~IX deals with the analysis of the limitations of our model. 
In Sec.~X we  evaluate   the GLD operation in the IR spectral range and  compare
 GLDs with some other photodetectors.
In Conclusions, we summarize the main results of the paper. 
The Appendix deals with the heat removal problem



\begin{figure}[htbp]
\centering
\includegraphics[width=5.0cm]{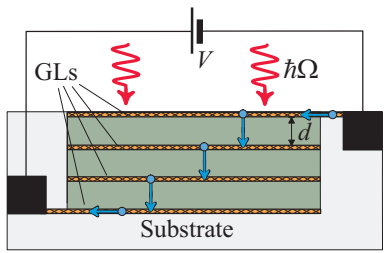}
\caption{Schematic structure  of vertical  GLDs  based on multiple-GL structure (with minimum of two GLs). The arrows show the current flow (for the case when all electrons crossing a GL are captured in it, i.e., for capture probability $p_c = 1$).
}
\end{figure}

\begin{figure*}[htbp]
\centering
\includegraphics[width=12.0cm]{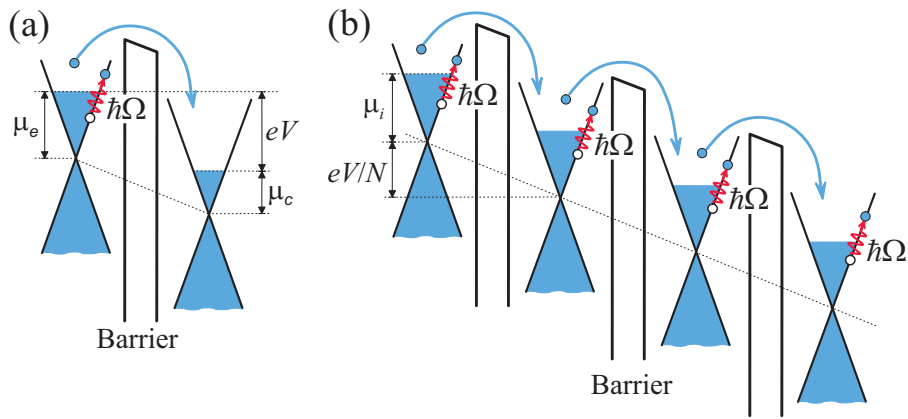}
\caption{Band diagrams of (a)   double-GLD and (b) 
multiple-GLD (with the cascade electron inter-GL transitions)  under applied bias.
The wavy arrows indicate the intraband (Drude) absorption, while smooth arrows correspond to thermionic emission processes resulting in the electron inter-GL transitions between neighboring GLs and providing  the dark current and photocurrent. The inter-GL transition between the distant GLs, which are possible at  finite values of the capture probability,  are not shown).}
\end{figure*}

\section{Device structures and principle of operation}

We consider two types of the  GLDs: (a) based on the n-doped  double-GL structure and (b) n-doped multiple-GL structure with the GLs separated by the barrier layers made of WS$_2$
or similar material with a relatively small conduction band off-set.  
As an example, Fig.~1 shows a GLD  using a four-GL structure.
The double-GLDs consist of only the top and bottom GLs  serving as the emitter and collector, respectively (no inner GLs).
In the multiple-GLDs, the inner GLs clad by the emitter and collector GLs 
are disconnected from the contacts.
In the double-GLDs (with a single barrier), the bias voltage $V$ applied  between the  top and bottom GLs 
induces the negative electron charge in   the emitter GL  the equal positive charge
in the collector  GL.
If the equilibrium electron concentration is low and the bias voltage is sufficiently strong,
the hole gas will be formed in the collector GL. 
In GLDs  with multiple-GL structures, the inner GLs remain  quasi-neutral,
so that the electron gas in each GL is formed primarily
due the n-type doping, whereas the top and bottom  GLs can be charged due to the bias voltage.
Figure~2 shows the GLPD band diagrams under the bias.
It is assumed that the GLDs under consideration are  irradiated by the normally  incident THz photons with the energy $\hbar\Omega$. 
The operation of GLDs is associated with the   electron 
heating due to the intraband absorption (Drude absorption) and the interband absorption
(see, for example,~\cite{32})
of the incident radiation resulting in an increase of the thermionic current over the barrier layers. 
Thus, 
the proposed GLDs are the barrier hot-electron bolometers.
In GLDs with the double-GL structures,
 the  electrons entering from the emitter GL and exiting to the collector GL support 
 the lateral current flowing via the contacts, so that the carrier densities in the GLs are maintained.
 In the multiple-GL structures,   
the  electron density in each GL between the emitter and collector  GLs  
is maintained due to the balance between the electrons leaving and entering GLs via the adjacent barriers  due to the thermal emission and the capture processes.
If the probability of the capture of an electron crossing a GL is smaller than unity, the GLD operation can exhibit the effect of photoelectric gain.
The origin of this gain  is of the same nature as in the vertical quantum-well infrared photodetectors (QWIPs)~\cite{33,34,35,36}.

The donor density $\Sigma_i$    and the bias voltage $V$
determine the
electron Fermi energies $\mu_e$ and $\mu_c$ in the top (emitter) and bottom (collector) GLs, respectively,
 ($\mu_e > \mu_h$, because the bias voltage increases the electron
density in the emitting GL and decreases it in the collecting GL). 
Considering the geometrical and quantum capacitances~\cite{37} and taking into account the energy gap between the Dirac points in GLs [see Fig.~2(a)] in the double-GL structure,
at relatively low bias voltages  one can obtain:

\begin{equation}\label{eq1}
\mu_e \simeq \mu_i\biggl[1 + \frac{eV}{2(eV_i + \mu_i)}\biggr], \,
\mu_c \simeq \mu_i\biggl[1 - \frac{eV}{2(eV_i + \mu_i)}\biggr],
\end{equation}
\begin{equation}\label{eq2}
\mu_i  =  \hbar\,v_W\sqrt{\pi\Sigma_i}
\end{equation}
Here $V_i = 4\pi\Sigma_ied/\kappa$, $e$ is the electron charge, $\hbar$ is the Planck constant,  $v_W \simeq 10^8$~cm/s
is the characteristic velocity of electrons and holes in GLs,and $\kappa$ and $d$ are the dielectric constant and the thickness of the barrier,
respectively.
At $\Sigma_i= (1.0 - 1.8)\times 10^{12}$~cm$^{-2}$, $\kappa = 4$ and $d  = 10 - 50$~nm,
one obtains $\mu_i \simeq 100 - 150$~meV and $V_i \simeq 452 - 3630$~mV. Relatively large values of $V_i$ imply that
for  the realistic moderate values of $V$ considered in the following, the correction of the Fermi energies
in the emitter and collector GLs is small in comparison with $\mu_i$. 
In the multiple-GL structures  (with a large number of GLs and the inter-GL barriers $N \gg 1$),
all the GLs except the top and bottom one's are quasi-neutral.
Although the electrically-induced variation of the Fermi energies in the emitter and collector GLs can be essential (for the mechanism of the photoelectric gain),
we will assume that in all GLs,  including the top and bottom one's,  the Fermi energies
are close to each other and approximately equal to the value determined by the donor density: 

\begin{equation}\label{eq3}
\mu \simeq \mu_i.
\end{equation}

\section{Vertical electron dark  current and  photocurrent}

 We restrict our consideration to  the double- and multiple-GL structures with relatively thick inter-GL barriers,
so that the tunneling current between the GLs can be neglected (the pertinent calculations can be done using the approach developed in Refs.~\cite{38,39}. We assume that the main contribution to
the vertical current is due to the thermoemission of  electrons resulting in the inter-GL transitions (producing the dark current). The impinging THz irradiation heats
the electron gas in GLs. This leads to
an increase in the thermoemission rate  intensifying of the inter-GL transitions, and, hence,
 the vertical current. 
The direct electron photoemission is insignificant  
 when the energy of photons $\hbar\Omega$ is smaller than the GL-barrier conduction band off-set $\Delta_C$ (the height of the barrier with respect of the Dirac point). 
 For the GL structures with the WS$_2$ barriers~\cite{40} it implies $\hbar\Omega < \Delta_C \simeq 0.4$~eV. Hence,  this inequality is well satisfied for the THz radiation.

 The rate of the thermionic emission from a GL (per unit of its area)  is given by
 \begin{equation}\label{eq4}
\Theta = 
\frac{\Sigma_i}{\tau_{esc}}\exp\biggl(\frac{\mu_i - \Delta_C}{k_BT}\biggr),
\end{equation}
 where  $T$ is the effective electron temperature which (under the irradiation) is higher than the lattice temperature
 $T_l$), $k_BT$ is the Boltzmann constant, and  $\tau_{esc}$ is the characteristic time of escape from the GLs of the  electrons with the energy $\varepsilon > \Delta_C$,
 $\tau_{esc} \sim \tau$, where $\tau$ is the momentum relaxation time.
Using  Eq.~(4) and  assuming for simplicity that $eV/N > k_BT$ ($V/N$ is the voltage drop across the barrier, and  taking into account  the electrons photoexcited  from the emitter 
and the photoexcited from  and captured to the  internal GLs (in multiple-GLDs), we find  the thermionic current density, $j$:

$$
j = \frac{e\Theta}{p_c}
= \frac{e\Sigma_i}{p_c\tau_{esc}}\exp\biggl(\frac{\mu_i - \Delta_C}{k_BT}\biggr)
$$
\begin{equation}\label{eq5}
\simeq \frac{e\mu_i^2}{\pi\hbar^2v_W^2p_c\tau_{esc}}\exp\biggl(\frac{\mu_i - \Delta_C}{k_BT}\biggr).
\end{equation}
 Here $p_c$ is the probability of the capture of an electron crossing a GL. 

In the  GL structures with at least one  internal GL (and in the multiple-GL structures),
the effects of the balance of 
thermogeneration from and capture to each GL, are taken into account by
 introducing the capture probability $p_c$,  as in 
 the standard models   of QWIPs~\cite{33,34,35,36}. In such an approach, the rate of the electron capture into each GL is equal to $p_cj/e$. 
Equating the capture rate  $p_cj/e$ and the  thermogeneration rate $\Theta$, one obtains
 $j = e\Theta/p_c$ [Eq.~(5)]. The quantity $p_c^{-1}$ can be relatively large
 if the capture probability is small. This quantity essentially determines the dark current and photocurrent gain $g \propto 1/p_c$.
 
Equation~(5) yields the following formula for the current density $j_0$ without irradiation
(i.e., for the dark current) when the dark electron temperature $T$ is equal to the lattice temperature
$T_0$: 
 \begin{equation}\label{eq6}
j_0 \simeq \frac{e\mu_i^2}{\pi\hbar^2v_W^2p_c\tau_{esc}}\exp\biggl(\frac{\mu_i - \Delta_C}{k_BT_0}\biggr).
\end{equation}
 
 In the double-GLDs
all the electrons generated by the emitter GL are captured by the collector GL, so that
in such a case $p_c = 1$.

Considering the variation of the electron temperature $T - T_0$, 
the photocurrent density $j - j_0$ 
can  be presented as

\begin{equation}\label{eq7}
j - j_0 = j_0\biggl(\frac{\Delta_C - \mu_i}{k_BT_0}\biggr)\frac{(T - T_0)}{T_0}.
\end{equation}

\section{Electron heating by incoming THz radiation}

As previously~\cite{22,32}, we assume that the electron energy relaxation is associated with
the processes of the emission and absorption of optical phonons.
In this case, for the rate, $\hbar\omega_0R$, of the energy transfer from the electron system to the optical phonon system is determined by (see, for example,~\cite{22,32}):

\begin{equation}\label{eq8}
R = \frac{\Sigma_i}{\tau_{0}}\biggl[({\cal N}_0 + 1)\exp\biggl(- \frac{\hbar\omega_0}{k_BT}\biggr) - {\cal N}_0\biggr]
\end{equation}
Here $\hbar\omega_0$ and ${\cal N}_0$ are the energy and the number of optical phonons, respectively, $\tau_0$ is the characteristics time of the optical phonon spontaneous emission  for the electron energy
 $\varepsilon > \hbar\omega_0$.

 If the characteristic time of the optical phonons decay $\tau_0^{decay} \ll \tau_0$,
${\cal N}_0$  is close to its equilibrium value:
 ${\cal N}_0 = [\exp(\hbar\omega_0/k_BT_0) - 1]^{-1} \simeq \exp(- \hbar\omega_0/k_BT_0)$.
 In the case of $\tau_0^{decay} >   \tau_0$, the effective energy relaxation time
 $\tau_0$ should be replaced by $\tau_0(1 + \xi_0)$ (where $\xi_0 = \tau_0^{decay}/\tau_0$)~\cite{14}.

When the effective electron temperature in GLs deviates from its equilibrium value (due to the absorption of THz radiation),
the  energy relaxation rate 
can be presented as [see Eq.~(8)]

\begin{equation}\label{eq9}
R \simeq \frac{\Sigma}{\tau_0}\biggl(\frac{\hbar\omega_0}{k_BT_0}\biggr)
\exp\biggl(- \frac{\hbar\omega_0}{k_BT_0}\biggr)
\frac{(T - T_0)}{T_0}.
\end{equation}

The rate of the energy transfer from the electron system to the optical phonon system $\hbar\omega_0R$ 
is equal to the rate, $\hbar\Omega G$, of the energy transferred from the THz radiation to  the electron system:

\begin{equation}\label{eq10}
\hbar\omega_0\ R = \hbar\Omega\,G. 
\end{equation}

Considering the intraband, i.e., the so-called free electron absorption (the Drude absorption)
and the interband absorption, the net absorption rate can approximately be presented as 
 
$$
G \simeq \beta\,I\biggl[\frac{D}{(1 + \Omega^2\tau^2)} 
$$
\begin{equation}\label{eq11}
+ \frac{\sinh(\hbar\Omega/2k_BT)}
{
\displaystyle\cosh(\hbar\Omega/2k_BT) + 
\displaystyle\cosh(\mu_i/k_BT)}\biggr]. 
\end{equation}
Here $\beta = \pi\,e^2/c_0\hbar \simeq 0.023$,  $c_0$ is the speed of light in vacuum, 
$I$ is the THz photon flux entering into the device (or the incident photon flux in the case of the anti-reflection coating),
and
\begin{equation}\label{eq12}
D = \frac{4k_BT\tau}{\pi\hbar}\ln\biggl[\exp\biggl(\frac{\mu_i}{k_BT}\biggr) + 1\biggr]
\simeq \frac{4\mu_i\tau}{\pi\hbar}
\end{equation}
is the  Drude weight,  the factor determining the  contribution of the Drude absorption
(it is proportional for the real part of the intraband conductivity of GLs).
For the  realistic values of $\tau$,
the factor $D$ can markedly exceed unity.
Indeed, assuming $\mu_i = 100- 150$~meV and $\tau = 10^{-13}$~s, one obtains
$D \simeq 20 - 30$. Strictly speaking, Eq.~(11) is valid at not too strong absorption. 

 Since the Fermi energy in the GLD under consideration should be sufficiently
large, the processes of the interband absorption of THz photons (their energy $\hbar\Omega \ll \mu_i$), 
corresponding to the second term in Eq.~(11), are effectively suppressed due to
the Pauli blocking. This implies that the electron heating by THz radiation is primarily
associated with the intraband absorption (with the Drude or the so-called free-electron absorption). 
In Eq.~(11) and in the following equations we  disregard
the attenuation in the multiple-GLDs of the THz photon flux associated with the absorption 
of in GLs, which are closer to the irradiated surface (emitter). This should be  valid at not too large values of $N$.

Taking into account
 the energy balance in each GL governed by Eq.~(10) and using Eq.~(11) (omitting the term
 describing the interband absorption), we arrive to the following expression for the variation of the effective electron energy caused by
 the THz of IR radiation of moderate intensity: 

\begin{equation}\label{eq13}
\frac{(T - T_0)}{T_0} = \frac{\beta\,D\tau_0(1 + \xi_0)I}{\Sigma_i(1 + \Omega^2\tau^2)} 
\biggl(\frac{k_BT_0}{\hbar\omega_0}\frac{\Omega}{\omega_0}\biggr)\exp\biggl(\frac{\hbar\omega_0}{k_BT_0}\biggr).
\end{equation}
Equation (13) corresponds to the electron energy relaxation time (determined by the optical phonons),
which is 
equal to~\cite{22}

\begin{equation}\label{eq14}
\tau_0^{\varepsilon} = \tau_0 (1 + \xi_0)
\biggl(\frac{k_BT_0}{\hbar\omega_0}\biggr)^2\exp\biggl(\frac{\hbar\omega_0}{k_BT_0}\biggr) \gg \tau_0.
\end{equation}

\section{Responsivity}

Using Eqs.~(6) and  (8), for the GLD responsivity  ${\cal R} = (j - j_0)/\hbar\Omega\,I$,   we obtain

$$ 
{\cal R} = \frac{e\mu_i^2}{\pi\hbar^2v_W^2p_c\tau_{esc}\hbar\Omega I}\biggl(\frac{\Delta_c - \mu_i}{k_BT_0}\biggr)
$$
\begin{equation}\label{eq15}
\times \exp\biggl(\frac{\mu_i - \Delta_C}{k_BT_0}\biggr)\frac{(T - T_0)}{T_0}.  
\end{equation}

Using  Eqs.~(13)  and (15), we arrive at the following expressions for the responsivity: 

\begin{equation}\label{eq16}
{\cal R} = \frac{\overline{ {\cal R}}}
{(1 + \Omega^2\tau^2)} 
\biggl(\frac{\mu_i}{\hbar\omega_0}\biggr)\biggl(\frac{\Delta_c - \mu_i}{\hbar\omega_0}\biggl)
\exp
\biggl(\frac{\mu_i + \hbar\omega_0 - \Delta_c}{k_BT_0}
\biggr)
\end{equation}
Here

\begin{equation}\label{eq17}
\overline{ {\cal R}}
=
\frac{4e\beta(1 + \xi_0)}{\pi\,p_c\hbar}
\biggl(\frac{\tau_0\tau}{\tau_{esc}}\biggr)
.
\end{equation}
As seen from Eq.~(16), the GLD responsivity is proportional to an exponential factor. To achieve reasonable GLD characteristics, the Fermi energy $\mu_i$ should not be too small in comparison with the barrier height $\Delta_C$. One can also see that ${\cal R} \propto {\overline{\cal R}} \propto 1/p_c$. As stated above, in the GLDs with the multiple-GL structures, the factor $1/p_c$ can be fairly large.

Equation~(16)  describes the GLD responsivity  as a function of the
THz radiation frequency $\Omega$, the temperature $T_0$ and the GL doping (via the dependence of $\mu_i$ on $\Sigma_i$).

Assuming $\hbar\omega_0 = 200$~meV, $\tau_0^{decay} + \tau_0 = 0.7$~ps,  $\tau_{esc}/\tau \sim 1.2$, and $p_c = 1$ for 
$T = 300$~K, from Eq.~(17)  we obtain from Eq.~(17)
$\overline{ {\cal R}} \simeq 27$~A/W.

Figure~3 shows the GLD responsivity versus the photon frequency $f = \Omega/2\pi$ calculated for different donor densities $\Sigma_i$ using Eqs.~(16)
and (17) for $\Delta_C = 400$~meV  and the same other  parameters as in the above estimate. 
This corresponds to the GLDs based on the double-GL structure
or to the GLDs based on  the multiple-GLDs  with a strong electron capture in the internal GLs.
The responsivity of the latter can be much higher than that shown in Fig.~3 if $p_c \ll 1$
(see below).

\begin{figure}[t]
\centering
\includegraphics[width=7.5cm]{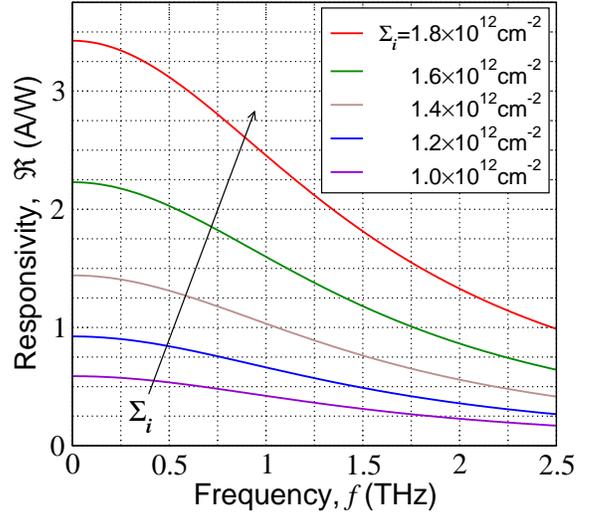}
\caption{Spectral dependences of  responsivity  of GLDs with different donor densities.
}
\end{figure}

\begin{figure}[t]
\centering
\includegraphics[width=7.5cm]{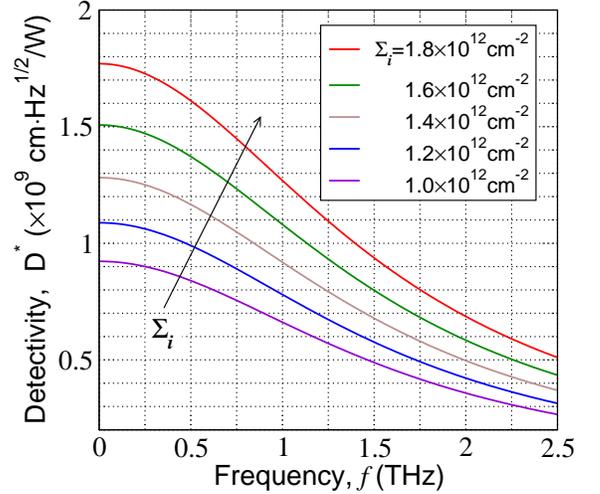}
\caption{Spectral dependences of detectivity of GLDs with  different donor densities 
 and $N/p_c = 25$).}
\end{figure}

\section{Dark current limited detectivity}

Considering that the shot noise current (at the value of the signal current equal to the dark current)  is given by $J_{noise} = \sqrt{4egJ_{dark}\Delta f}$,
where $\Delta f$ is the bandwidth and $g = 1/Np_c$ is the dark current and photoelectric gain,
the dark current limited detectivity  (see, for example, Ref.~\cite{36}), 
can be presented in the following form:

\begin{equation}\label{eq18}
 D^{*}
=\frac{{\cal R}}{\sqrt{4egj_0}}.
\end{equation}

Accounting for Eq.~(16), 
we arrive at

$$
D^{*} = \frac{{\overline D^*}}{(1 + \Omega^2\tau^2)} 
\biggl(\frac{\Delta_C - \mu_i}{\hbar\omega_0}\biggl)
$$
\begin{equation}\label{eq19}
\times\exp
\biggl(\frac{\mu_i - \Delta_C}{2k_BT_0}\biggr)
\exp
\biggl(\frac{\hbar\omega_0}{k_BT_0}\biggr)\sqrt{\frac{N}{p_c}},
\end{equation}
where

\begin{equation}\label{eq20}
{\overline D^*} = 2\sqrt{\pi}\beta
\biggl(\frac{k_BT_0}{\hbar\omega_0}\biggr)
\biggl[\frac{(1 + \xi_0)\tau_0\tau\,v_W}{\hbar\omega_0\sqrt{\tau_{esc}}}\biggr].
\end{equation}
For $\tau_0^{decay} + \tau_0 = 0.7$~ps,  $\tau \sim 0.1$, ps, $\tau_{esc} \sim 0.12$~ps, and
$T = 300$~K, 
$\Delta_C = 400$~meV, $\Sigma_i = 1.8\times 10^{12}$~cm$^{-2}$ ($\mu_i = 150$~meV), 
$N/p_c = 1 -  25$, and $f = \Omega/2\pi \ll 1.6$~THz from Eqs.~(19) and (20) we obtain 
 ${\overline D^{*}} \simeq 1.3\times 10^7$cm Hz$^{1/2}$/W and 
$D^{*} \simeq (0.35 - 1.75)\times 10^9$~cm Hz$^{1/2}$/W.
Figure~4 shows the spectral characteristics of GLDs with $\Sigma_i = 1.0\times 10^{12} - 1.8\times 10^{12}$~cm$^{-2}$ ($\mu_i \simeq 100 - 150$~meV) calculated using Eqs.~(19) and (20)
for the same other parameters as from the latter estimate and  Fig.~3.

From Eqs.~(16), (17), (19), and (20), one can see that the GLD responsivity is independent on $N$
(in the framework of the present model),
whereas the GLD detectivity is proportional to $\sqrt{N}$ (as in QWIPs~\cite{36}).

\section{Role of the electron capture}

 \begin{figure}[t]
\centering
\includegraphics[width=7.5cm]{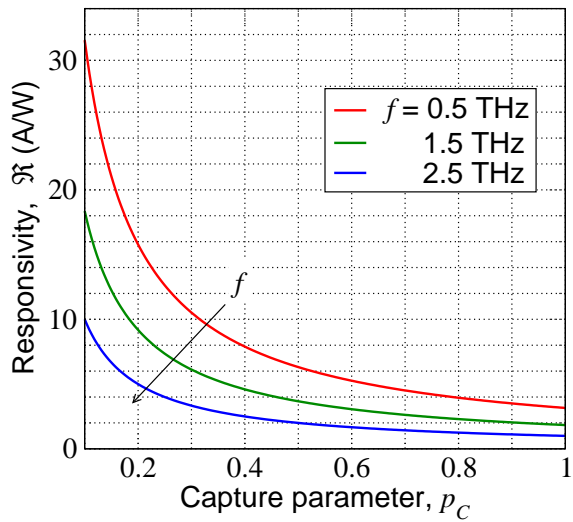}
\caption{Responsivity of GLD as a function of   the capture parameter $p_c$
for different radiation frequencies ($\Sigma_i = 1.8\times 10^{12}$~cm$^{-2}$).
}
\end{figure}

 \begin{figure}[t]
\centering
\includegraphics[width=7.5cm]{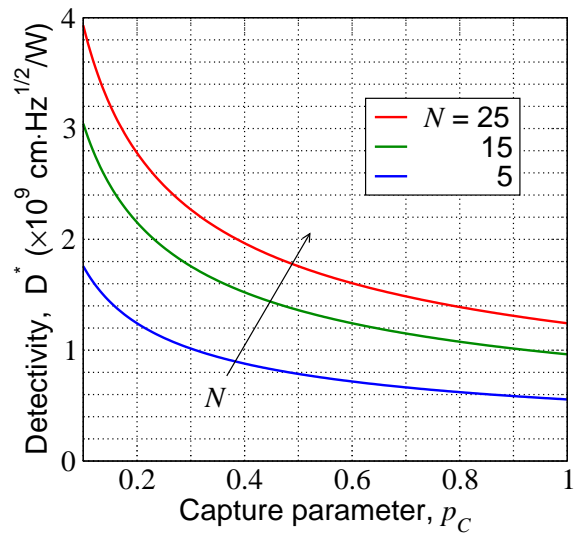}
\caption{Detectivity as a function of   the capture parameter $p_c$
for GLDs with different  number of the inter-GL barriers $N$ ($\Sigma_i = 1.8\times10^{12}$~cm$^{-2}$ and $f = 1$~THz).}
\end{figure}

As follows from Eqs.~(16), (17), (19), and (20), both the responsivity and detectivity
of the multiple GLDs increase with decreasing capture probability $p_c$, i.e., with increasing photoelectric gain. The latter quantity
is determined by several factors, in particular, by the degree of the electron heating 
in the inter-GL barriers and, hence, by the potential drop across these barriers and their thickness. The detailed calculations of $p_c$ require additional quantum-mechanical calculations of the electron transitions from the continuum states above the barriers to
the bound states in GLs coupled with the ensemble Monte Carlo modeling of the electron propagation
across the GL-structure similar to that made previously for  multiple-QW structures based on
the standard semiconductor heterostructures (see, for example, Refs.~\cite{35,41,42}).
This is, however, beyond the scope of this work, so that here we consider $p_s$ as a phenomenological parameter. Figures~5 and 6  show the GLD responsivity and detectivity as functions of the capture parameter. One can see that a decrease in the capture parameter $p_c$ leads to a substantial rise of ${\cal R}$. 
At low $p_c$, the GLD responsivity can be fairly high.
The detectivity   $D^*$ of GLDs with the multiple-GL structure also rises with decreasing $p_c$ as well as
with increasing $N$.
Since the capture probability $p_c$ in the multiple-GL structures should  markedly decrease with increasing electric field
in the barrier layers $E = V/Nd$ (as in multiple-QWIPs~\cite{35}),
the GLD responsivity and detectivity can be rising functions of the bias voltage
if the proper heat removal is provided.

\section{Effect of plasmonic resonances}

 Since the absorption of the incident THz radiation is associated with the Drude mechanism,
 the absorption efficiency and, hence, the GLD responsivity and detectivity can be relatively small in the frequency range $\Omega/2\pi > \tau^{-1}$. If $\tau \simeq 0.1$~ps,
 this corresponds to $\Omega/2\pi > 1.6$~THz. However, the operation of GLDs can be extended
 to  much higher frequencies if the GLD structure is supplied by a metal grating over the top GL (not shown in Fig.~1). In this case, the incident THz radiation can result in an efficient excitation
 of plasma oscillations  in the electron-hole system in the double-GL GLDs and
 in the  system of electrons in all GLs (in the multiple-GL structures).
 Simplifying  the equations from Ref.~{43}
 for the GLDs with a metal grating, the rate of the THz radiation absorption $G_n$ 
 at the frequency near the $n$-th plasmon resonance ($\Omega \simeq \Omega_n$) can be presented 
 as ~\cite{43}:
 
\begin{equation}\label{eq21}
G_n  = \frac{\beta\,I DA_n}{1 + (\Omega - \Omega_n)^2\tau^2a_n^2}. 
\end{equation}
Here  $A_n = 1/(1 + \beta\,D/2\sqrt{\kappa})^2 \simeq 1$ and $a_n = 4/(1 + \beta\,D/2\sqrt{\kappa})^2 \simeq 4$ are determined by the ratio of the collisional damping (which is  actually close to $1/2\tau$) and the parameter of the radiative damping~\cite{44}.
 Equation~(21) does not contain any geometrical parameters such as the grating period,
length of grating strips, and the spacing between the grating and the top GL.
These parameters only determine 
the dependence of the resonant plasma frequencies $\Omega_n$ on the device geometry. 
This is valid as long as those dimensions are much shorter than the THz radiation wavelength
and the net length of the grating is of the same order of magnitude as the wavelength.

The quantities $\Omega_n$  depend on the net electron density in all GLs $(N + 1)\Sigma_i$,  the spacing between the top GL and the metal grating $W$, and the period of the grating.
The latter determines the "quantized"
wave-number $q_n$ of the excited plasma modes (standing plasma waves). One can put  
$q_n = (\pi/2L)n$,  $2L$, the length of the GL-structure in the lateral direction,and $n = 1,2,3,...$ is the plasma mode index.
For simplicity, 
one can use the following equation for the frequency of the plasma modes (corresponding to
$q_nW \gtrsim 1$):
  
\begin{equation}\label{eq22}
\Omega_n \sim \sqrt{\frac{e^2\mu_i(N + 1)}{\kappa\hbar^2}q_n}, 
\end{equation}
or
  
\begin{equation}\label{eq23}
\Omega_n \sim \sqrt{\frac{\pi\,e^2\mu_i(N + 1)}{2L\kappa\hbar^2}n}. 
\end{equation}
The square-root dependence of $\Omega_n$ on $N$ appears because the net electron density,
which  determines the contribution to the self-consistent
electric field in the plasma waves  by all the GLs
is proportional to $(N + 1)$, whereas the electron fictitious mass $m_f$
in  GLs is proportional to $\mu_i \propto \sqrt{\Sigma_i}$  (see, for example, Ref.~\cite{39}).
Setting $\mu_i = 150$~meV, $2L/n = 0.5 - 1.0~\mu$m (i.e. $2L = 10~\mu$m and $n = 10$), and $N = 5$, from  Eq.~(23) we obtain $f_{10} = \Omega_{10}/2\pi \simeq 7.4 - 10.4$~THz. If $A_n \sim 1 $
and $a_n \sim 1$, the GLD responsivity at the resonance is of the same order of magnitude as
at the low edge of  the THZ range $\Omega \ll \tau^{-1}$ (see Figs.~3 and 5).
Thus,  
the resonant excitation of plasma oscillations results in 
 a strong absorption of the incident THz
 radiation  and, hence, in  elevated values of the GLD responsivity (and detectivity)
 at relatively high frequencies (several THz). Such GLDs can cover the frequency range  $f \simeq 6 - 10$~THz ($\hbar\Omega \simeq 25 - 40$~meV), which is not
 accessible by A$_3$B$_5$-based detectors, in particular, THz quantum-well detectors (QWDs)~\cite{46,47,48}.
 
\section{Limitations of the model}

The model used in the above calculations some simplifications.
These simplifications are:
(i) The capture probability is the same for all GLs in the GL-structures;
(ii) The thermoassisted tunneling is insignificant; 
(iii) The heating the Joule heating of the structure.

Since the capture probability $p_c$ depends on the heating of electrons in the barriers,
it can be determined on only by the average electric field in the GL-structure
but partially by the electric field in the adjacent barriers. In this case, the probability
of the electron capture to the particular GL can depend on its index. Such kind of non-locality
of the electric-field dependence can lead to more nontrivial spatial distributions (as in QWIPs~\cite{41,42}). However, in the GL-structures with the barrier thickness much smaller
than the characteristic energy relaxation length, the pertinent effect should be weak.
This justifies the assumption that $p_c$ is a constant (which  generally depends on the average electric field).

 At sufficiently  high bias voltages (much higher than those assumed above), the electron escape from GLs can be associated with the thermoassisted tunneling from the bound states in GLs to the continuum states above the barriers. This tunneling can also be used in double- and multiple-GLDs with
the structures similar to those considered above. Since the effective activation energy
for this mechanism can be markedly smaller that $(\Delta_C - \mu)$, GLDs with the thermoassisted 
tunneling can comprise  the barriers with larger conduction band offsets than between GLs and WS$
_2$, for example, with the $hBN$ barriers. However, this problem requires a separate consideration.

Above we considered the case of not too low bias voltages ($eV/N > k_BT$).
The Joule power $j_0V$ can result in an overheating of the GL structure if $V$ is relatively strong. Such an overheating can be avoided either by decreasing $\mu_i$
(decreasing the GL doping level) or by lowering the bias voltage $V$.
In the range of bias voltages $eV/N < k_BT$, the GLD responsivity and detectivity
given by Eqs.~(16) and (19) should be multiplied by the factors 
$\zeta =\{1 - \exp[-(eV/Nk_BT)]\} \simeq eV/Nk_BT$  and 
$\sqrt{\zeta} = \sqrt{1 - \exp[-(eV/Nk_BT)]}
\simeq \sqrt{eV/Nk_BT}$,  respectively. The transfer to the range of relatively low bias voltages leads to a decrease in the Joule power as $V^2$, but at the expense of a decrease in
the responsivity and detectivity (${\cal R} \propto V/N$ and $D^* \propto \sqrt{V/N}$).

The Joule heating can lead to overheating of GLDs if the Joule power exceeds the maximum
heat energy which can be removed from the GLD unit area, $W_{max}$, without a substantial heating.
This results in the following limitation:

\begin{equation}\label{eq24}
W^{max} > j_0VA = \frac{e\Sigma_i}{p_c\tau_{esc}}\exp\biggl(\frac{\mu_i - \Delta_C}{k_BT_0}\biggr)V, 
\end{equation}
where $A$ is the device area.
Assuming a typical voltage drop cross the GL-structure to be on the order of 50 - 500~mV and the thermal resistance of the package to be on the order of 10 K/W, we obtain that $W^{max}$ and the current leading to the ten degrees overheating $j_0^{max}$ are equal to 1~W and  2 - 20~A, respectively. For a typical 300$\times 300~\mu$m$^2$ device, this corresponds to a fairly reasonable current density of $j_0^{max}\sim 2\times (10^3 - 10^4)$~A/cm$^2$. 
Setting $\Sigma_i = 2\times 10^{12}$~cm$^{-2}$, $\tau_{esc} = 0.1$~ps, and $p_c = 0.5$, 
we obtain $j_0 \sim 3\times 10^2$~A/cm$^2$ (i.e., $j_0 < j_0^{max}$.
Much higher current densities could be achieved with improved heat sinks (see, for example, Ref.~\cite{49}) and/or in the pulsed regime of operation.

\section{Discussion}

tunneling can be based on the materials with larger conduction band offsets than between GLs and WS$_2$.  


In principle,  GLDs  can also effectively operate in the mid- and near-IR ranges. At sufficiently high photon energies, the intraband absorption is negligible, whereas the interband
radiative processes, corresponding to the second term in the right-hand side of Eq.~(11), can efficiently contribute to the heating of the electron gas in GLs if $\hbar\Omega \gtrsim 2\mu_i$.
In such a case  for the  photon energies $2\mu_i < \hbar\Omega < 2\Delta_C$,
the GLD responsivity is given by 

\begin{equation}\label{eq25}
{\cal R}_{IR} \simeq {\tilde{ {\cal R}}}
\biggl(\frac{\Delta_c - \mu_i}{\hbar\omega_0}\biggl)
\exp
\biggl(\frac{\mu_i + \hbar\omega_0 - \Delta_C}{k_BT_0}
\biggr),
\end{equation}

\begin{equation}\label{eq26}
\tilde{ {\cal R}}_{IR} = \frac{\pi\beta\,e(1 + \xi_0)}{p_c\hbar\omega_0}\biggl(\frac{\tau_0}{\tau_{esc}}\biggr)
\biggl(\frac{k_BT_0}{\hbar\omega_0}\biggr).
\end{equation}
At $\Sigma_i = (1.0 - 1.8)\times 10^{12}$~cm$^{-2}$ ($\mu_i \simeq 100 - 150$~meV),
Eqs.~(25) and (26) yield the values of the responsivity ${\cal R}_{IR} $ about 20-30 times smaller than ${\cal R}$
in the range $\Omega \ll 1/\tau$ (see Figs.~3 and 5). In particular, at  $\Sigma_i =  1.8\times 10^{12}$~cm$^{-2}$ , assuming  $p_c = 0.2 - 1.0$, we obtain rather high values ${\cal R}_{IR} \simeq 0.11 - 0.55$~A/W.  The GLD detectivity in the mid- and near-IR range $D^*_{IR}$, being much lower
than $D^*$ in the THz range,  can be still relatively high (for  room temperature).
Note that $\tilde{ {\cal R}}_{IR}$ are $D^*_{IR}$ independent of the photon energy in its wide
range (from $200 -300$~meV to 800~meV).


Comparing the  GLDs based on the vertical double-GL structure under consideration with the GLDs with a lateral structure and the barrier region consisting of  an array of graphene nanoribbons using the electron heating in n-GL contact region,~\cite{22} one can see that both types of THz detectors at the room temperature exhibit close spectral characteristics. However, the GLDs with the vertical
multiple-GL structure can have much higher responsivity and, especially, detectivity
if $p_c < 1$ and $N \gg 1$.

In principle, room-temperature THz detectors utilizing the thermionic emission of electrons heated by the absorbed THz radiation from QWs can be made of A$_3$B$_5$ or  Si-Ge heterostructures.
Such detectors on the base of vertical multiple-QW structures were proposed and realized a long time ago(see Refs.~\cite{50} and~\cite{51}, respectively, as well as a recent paper~\cite{52}). The THz detectors based on lateral
structures with the barrier regions formed by the metal gates were also realized~\cite{53,54}
(see also Ref.~\cite{55}).
However, the responsivity and detectivity of  GLDs under consideration can be markedly higher
than that using  the A$_3$B$_5$ multiple-QW structures.
Comparing the Drude factor $D$  for GL-structures
[see Eq.~(12)] and the same factor $D_{QW}$ for QW-structures with GaAs QWs, one can find the ratio
of these factors at the equal electron density $\Sigma_i$ and momentum relaxation time $\tau$ is given by
\begin{equation}\label{eq27}
\frac{D}{D_{QW}} \simeq \frac{mv_W^2}{\mu_i} \simeq \frac{m}{m_{f}}, 
\end{equation}
where $m$ and $m_f$ are the effective and fictitious  electron masses in QWs and GLs, respectively. For GaAs QWs and GLs with $\mu_i \simeq 150$~meV, these masses are approximately equal to each other. This implies that the  THz  power absorbed in QWs and GLs are close.
However, the electron energy relaxation time in GLs is longer than that in GaAs-QWs  and other
standard semiconductor QWs.
This is mainly due to relatively large optical phonon energy in GLs.
Indeed, using Eq.~(14) and  assuming that $\tau_0^{decay} + \tau_0 = (0.7 -1.4)$~ps at the room temperature we obtain
$\tau_0^{\varepsilon} \simeq (32.5 -65)$~ps, while for GaAs ($\hbar\omega_0 \simeq 36$~meV and $\tau_0 \simeq 0.14$~ps), InAs ($\hbar\omega_0 \simeq 30$~meV and $\tau_0 \simeq 0.2$~ps), and InSb ($\hbar\omega_0 \simeq 25$~meV and $\tau_0 \simeq 0.7$~ps) QWs one obtains
$\tau_0^{\varepsilon} \simeq 0.56$, 0.93 and 3.93~ps, respectively.
Longer electron energy relaxation time corresponds to more effective heating of the electron gas
and, hence, higher responsivity. An other factor promoting higher responsivity (and detectivity)
of GLDs is the possibility to achieve higher photoelectric gain due to smaller values of the expected capture parameter $p_c$.

The THz QWPs using the direct intersubband photoexcitation from QWs require the heterostructures
with rather small band off-sets ($\Delta_C \sim \hbar\Omega$). They exhibit a modest responsivity (about few tens of mA/W or less~\cite{46,47,48}) with $D^* \simeq 5\times 10^7$
cm Hz$^{1/2}$/W at $T_0 = 10$~K~\cite{46}.
Hence, in the few-THz range, GLDs surpass QWPs. GLDs with the grating using the plasmonic effects although should exhibit advantages over  QWPs in the range 6 - 10 THz (see above).
Additional advantages of GLDs might be associated with better heat removal conditions~\cite{49,56,57} than
in the case of different A$_3$B$_5$ devices.
  

Due to a substantial progress in fabrication and experimental studies of the multiple-GL structures
with the inter-GL barrier layers made of transition metal dichalcogenides~\cite{19} (see also Refs.~\cite{58,59,60,61,62}), 
the realization of  the proposed GLDs appears to be feasible. In particular, similar GL-structures
with five periods and 20~nm thick barriers~\cite{58} and with ten periods~\cite{59}
were demonstrated.

\section{Conclusions}

We proposed THz GLDs based on the double-GL  and  multiple-GL structures with the barrier layers made of WS$_2$ exploiting the enhanced thermionic electron emission from GLs due to the intraband
(Drude) absorption, developed the device model, and calculate the GLD responsivity and detectivity
at the room temperature. We demonstrated that GLDs, especially, those based on the multiple-GL structures can exhibit fairly high responsivity and detectivity surpassing hot-electron detectors 
based on the standard heterostructures. The main advantages of GLDs are associated with 
relatively long electron energy relaxation time and the pronounced effect of  photoelectric gain
at a  low capture probability of the electron capture into GLs.
As shown, GLDs using the resonant electron heating associated with the plasmonic effects
and GLDs exploiting the electron heating due to the  interband absorption
  can also operate in the far-, mid, and near-IR ranges of the radiation spectrum.

\section*{Acknowledgments}

This work was supported by the Japan Society for Promotion of Science (Grant-in-Aid for Specially Promoting Research $\# 23000008$),  Japan. V. R. and M. R. acknowledge  the support
of the Russian Scientific Foundation( Project $\# 14-29-00277$).
The work at the University at Buffalo was supported by the NSF  TERANO grant and the  US Air Force Office
of Scientific Research.
 The work at RPI was supported by the US Army
Cooperative Research Agreement.


\end{document}